\documentclass{iaus}
\usepackage{graphicx}
\usepackage{mathastext}

\title[~~Superthin LSB galaxy UGC 12281] 
{On the nature of the peculiar\\superthin LSB galaxy UGC 12281}

\author[Philip G{\"u}nster \& Dominik J. Bomans]   
{Philip G{\"u}nster
 \and Dominik J. Bomans}

\affiliation{Astronomisches Institut, Ruhr-Universit{\"a}t Bochum, \\ Universit{\"a}sstra\ss e 150,
D-44801 Bochum, Germany \\ email: {\tt guenster@astro.rub.de}, {\tt bomans@astro.rub.de}}

\pubyear{2011} 
\volume{284}  
\pagerange{1--12}
\jname{The Spectral Energy Distribution of Galaxies}
\editors{R.J. Tuffs \&  C.C.Popescu, eds.}
\begin{document}

\maketitle

\begin{abstract}
UGC 12281 has been classified as having a pure disk and being a low surface brightness galaxy (LSBG), thus being an obvious member of the so-called superthin galaxies. At the same time it represents an extremely untypical type of LSBG due to its remarkable amount of current star formation and evidence for extraplanar ionized gas. This makes it become a perfect tool to investigate the triggering of star formation in LSB galaxies, located in an alleged isolated area. By means of deep photometry and long-slit spectroscopy we analyse the H$\alpha$ halo and verify the existence of a potential dwarf companion which we found on processed SDSS images.

\keywords{galaxies: evolution, galaxies: dwarf, galaxies: halos}
\end{abstract}

\firstsection 
\section{Introduction}
\noindent Galaxies in their variety reveal an interesting subset, very late-type spirals without bulge viewed edge-on. They were first mentioned by \cite{GoadRoberts81} and called superthin galaxies from then on. The superthins can be distinguished easily as having very thin disks causing a measured axis ratio of more than 9. The modest gradients of their rotation curves to the center imply a small central mass concentration. This is in good agreement with their tendency to not show any central bulge. Based on the edge-on disk galaxy catalog  by \cite{Kautsch06} about 38$\%$ among the superthins are LSB galaxies showing a blue central surface brightness of more than 23\,mag$\cdot$arcsec$^{-2}$ (\cite{ImpeyBothun97}). This implies they must have a relatively small star formation rate (SFR) over large time scales. To address the question why they never formed enough stars to appear significantly brighter, \cite{Rosenbaum09} looked at the large-scale environment in which they are embedded. They conclude that LSBGs must have evolved in low density areas without any tidal interaction with companions to effectively trigger star formation. Therefore superthin LSB galaxies with high SFR promise to be an excellent tool to understand LSBG evolution. We investigate here the superthin UGC 12281. This galaxy has an axis ratio of almost 12. However, \cite{RossaDettmar03} revealed it as H$\alpha$ bright implying a high star formation activity. Indeed they determined a radial extent of strong star formation to 20\,kpc which accounts for 2/3 of the total disk. Hints for extraplanar diffuse ionized gas (eDIG) were detected, too.   

\section{Data}
\begin{figure}[htb]
\begin{minipage}[b]{0.49\textwidth}
\centering\includegraphics[height=4.5cm]{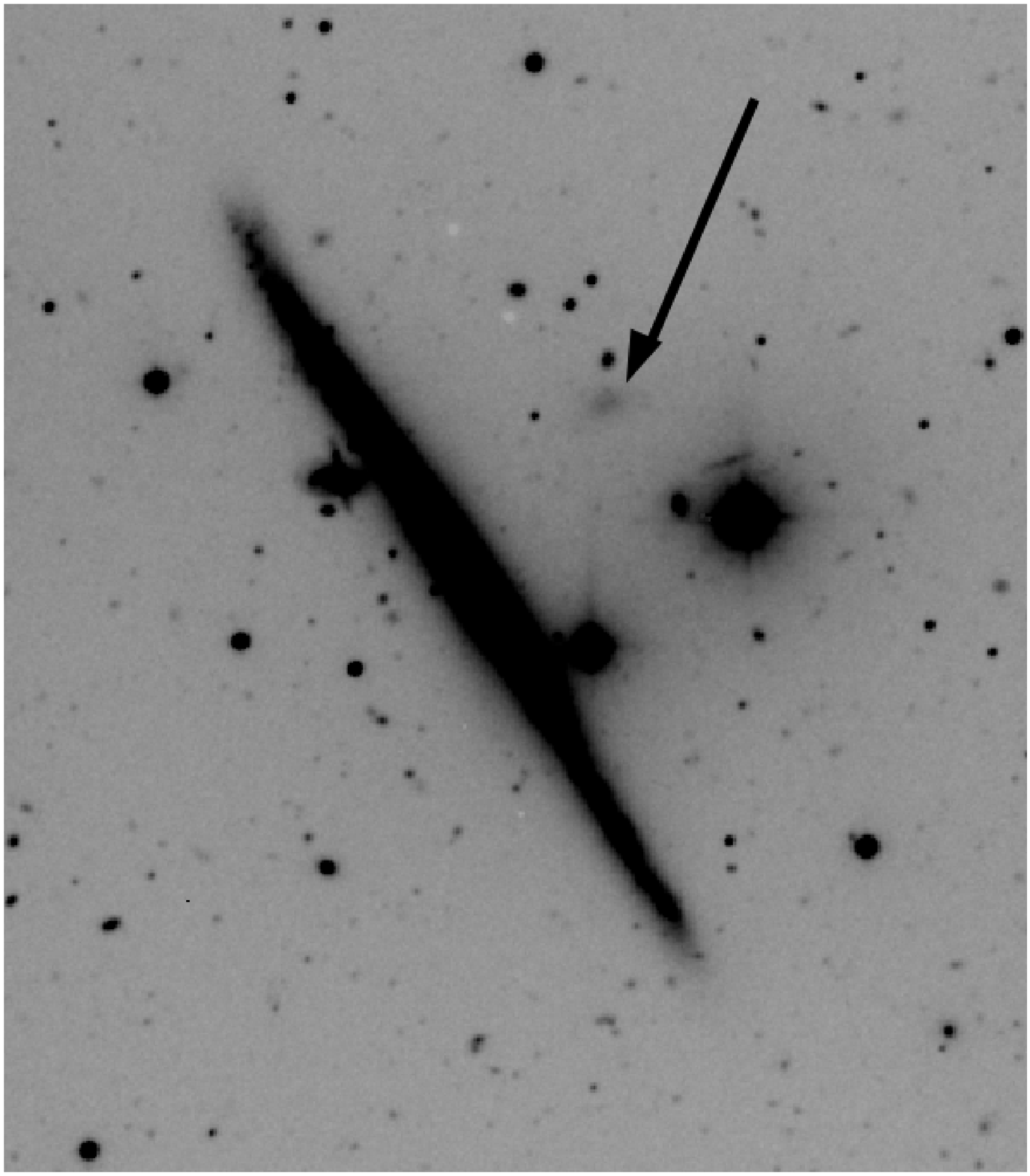}
\caption{Roeser BV image of UGC 12281 taken with the CAHA 2.2\,m telescope. The arrow marks the position of the dwarf companion.}
\label{fig1}
\end{minipage}
\hfill
\begin{minipage}[b]{0.49\textwidth}
\centering\includegraphics[height=4cm]{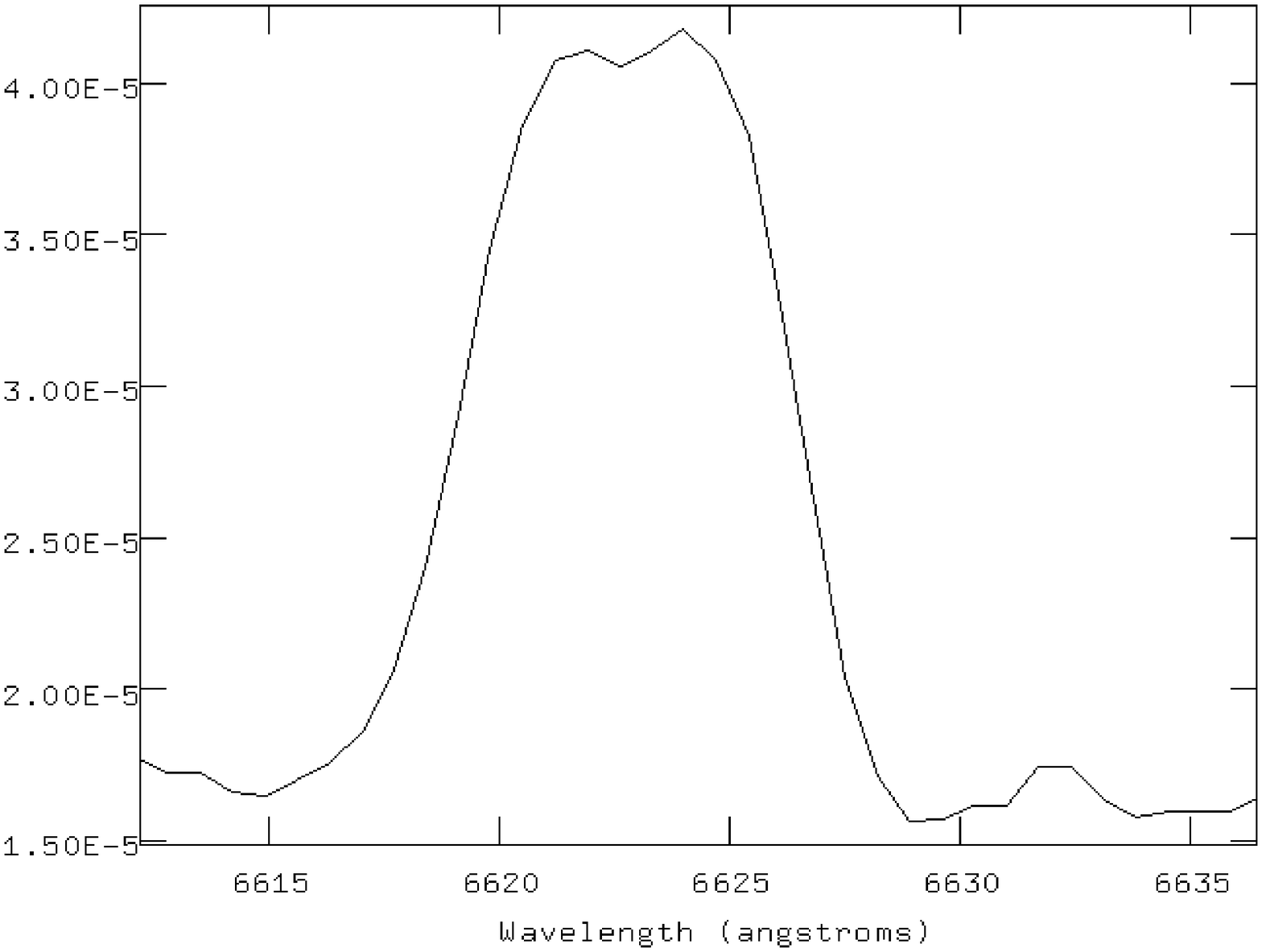}
\caption{Closeup view of the H$\alpha$ line in the disk spectrum of UGC 12281. The line obviously has a double peak. The split was measured to 170\,km$\cdot$s$^{-1}$.}
\label{fig2}
\end{minipage}
\end{figure}

We observed UGC 12281 with the 2.2\,m telescope at the Calar Alto Observatory (CAHA). The CAFOS instrument provided deep imaging data in the bands Roeser BV, R and H$\alpha$. We also did long-slit spectroscopy at two slit positions: first we aim at measuring an integral spectrum of the diffuse halo gas. For that the slit is placed parallel to the disk with a 2\,kpc offset from the midplane. Secondly, a slit is set perpendicular to the galaxy's major axis hitting the spot of the suspected dwarf companion (position flagged with an arrow in Fig. \ref{fig1}) and crossing the disk. 

\section{UGC 12281}
Our structural analysis includes disk profile fits: We trace its shape first using the Source Extractor software by \cite{BertinArnouts96}. The derived astrometric and photometric parameters of the Roeser BV image are fed into the Galaxy Fitting code GALFIT (\cite{Peng10}). We apply a two-dimensional single profile to the disk according to \cite{Sersic68}, namely of the kind I(R)\,=\,I$_0\cdot$ exp$\left[-\left(\frac{R}{a}\right)^{1/n}\right]$ where I$_0$ is the central intensity and a the scale length. The S\'ersic index n is found to 1.25 which indicates pure-disk structure. GALFIT also provides the residual of the original and the model fit which shows no sign of a bulge, but spiral structure.

\section{A close-by dwarf companion}
From archival data of the Sloan Digital Sky Survey (SDSS) we were able to derive the following parameters of the possible dwarf companion: Its absolute magnitude is M$_{g'}$\,=\,-12.01 and the average surface brightness $\mu_g'$\,=\,25.73\,mag$\cdot$arcsec$^{-2}$. The size could be estimated to approx. 2.0\,x\,1.0\,kpc. The dwarf is elongated towards the host galaxy hinting at gravitational interaction with UGC 12281. If truly a satellite galaxy, this may be part of the explanation for UGC 12281's relatively high star formation providing stirring of the disk. Compared to the Milky Way satellite Fornax (M$_V$\,=\,-13.3, major axis\,=\,2.8\,kpc, $\mu_g'$\,=\,23.4\,mag$\cdot$arcsec$^{-2}$; \cite{Lokas09}), the companion is similar in absolute magnitude, but more compact and with lower surface brightness, again supporting the idea of gravitational disturbance. Its B-V color is measured to 1.13 which is redder than Fornax (B-V\,=\,0.63; \cite{Mateo98}) and implies a higher metallicity or an intermediate-age stellar population. 

\section{A nearly transparent disk?}
The slit orientation parallel to the galaxy's disk but slightly above it should reveal the diffuse halo gas. Unfortunately, in our initial spectral analysis no emission line features are detected while deep H$\alpha$ imaging shows diffuse H$\alpha$ halo emission. When we put the slit going through the galaxy's disk the H$\alpha$ and the [N\,II] line appear to have double peaks (visible in Fig. \ref{fig2}) and account for a split of 170\,km$\cdot$s$^{-1}$, similar to the HI v$_{rot,max}$\,=\,146\,km$\cdot$s$^{-1}$ (\cite{Warmels88}). This raises the question if we detected both the fore and the back side of the rotating disk. Since we look at an extreme flat edge-on object this assumption would require UGC 12281 to be nearly complete transparent. A multi-color analysis shows neither in the SDSS nor in our deep data an obvious dust lane. Such a smooth distribution of dust components is in agreement with results for other LSBGs of MacLachlan et al. (this conference).  

\section{Chain galaxies}
Superthin galaxies with strong star formation may be in close connection to objects called chain galaxies. They are high axis ratio, clumpy systems at high redshifts. Although edge-on galaxies have obviously more extinction than the ones viewed face-on, their average surface brightnesses can be brighter (\cite{Elmegreen04}). This is due to their extinction path length which is usually larger than the actual disk thickness. \cite{Holmberg58} showed that the inclination-corrected face-on surface brightness is brighter than the actual average surface brightness by the extinction effect. This means that this distant galaxies can compensate for the effect of the cosmological Tolman dimming. The high SFR superthins, in the end, may be the low-redshift counterparts of chain galaxies.

\section{Conclusions \& Outlook}
UGC 12281 appears to be an untypical LSB galaxy. Its recent star formation is significant and we found strong hints for a diffuse H$\alpha$ halo. The disk is well fitted by a S\'ersic index of 1.25 with no strong disturbances. The H$\alpha$ line shows a split which is about in the order of the rotational velocity. This implies a nearly transparent disk. Our data verifies the existence of a very low surface brightness dwarf companion which is a candidate for interacting with its host and thus triggering its star formation. Star forming LSB superthins like UGC 12281 could allow us to overcome the Tolman dimming and extend our study of LSBG evolution to much higher redshifts.

\end{document}